\documentclass[trackchanges,twocolumn,twocolappendix]{aastex701}

\usepackage{multirow}
\usepackage{makecell}
\usepackage{amsmath}

\newcommand{\Nii}{[\rm N\textsc{ii}]}
\newcommand{\NiiHa}{[\rm N\textsc{ii}]_{6584}/H\alpha}
\newcommand{\OiiiHb}{[\rm O\textsc{iii}]_{5007}/H\beta}

\begin{document}

\title{Fast Radio Bursts Associated with Persistent Radio Sources Prefer Hydrogen-poor Superluminous Supernova Host Environments}

\correspondingauthor{Fujia Li, Di Li}

\author[0000-0001-9472-2052]{Fujia Li}\email[show]{lifujia0928@gmail.com}
\affiliation{Department of Astronomy, Tsinghua University, Beijing 100084, China}

\author[0009-0003-8087-2991]{Fei Liu}\email{f-liu25@mails.tsinghua.edu.cn}
\affiliation{Department of Astronomy, Tsinghua University, Beijing 100084, China}

\author[0000-0003-3010-7661]{Di Li}\email[show]{dili@mail.tsinghua.edu.cn}
\affiliation{New Cornerstone Science Laboratory, Department of Astronomy, Tsinghua University, Beijing 100084, People’s Republic of China}
\affiliation{Department of Astronomy, Tsinghua University, Beijing 100084, China}

\author[0000-0001-7931-0607]{Dongzi Li}\email{dzli@tsinghua.edu.cn}
\affiliation{Department of Astronomy, Tsinghua University, Beijing 100084, China}

\author[0009-0006-4677-3780]{Ziheng Ding}\email{ding-zh25@mails.tsinghua.edu.cn}
\affiliation{Department of Astronomy, Tsinghua University, Beijing 100084, China}

\author[0009-0009-3255-4132]{Ran Gao}\email{gaor@mail.tsinghua.edu.cn}
\affiliation{Department of Astronomy, Tsinghua University, Beijing 100084, China}

\begin{abstract}
A subset of repeating fast radio bursts (FRBs) is associated with compact persistent radio sources (PRSs).
The FRB-PRS hosts differ dramatically from the general FRB host population: they do not follow the distribution of localized non-PRS FRB hosts, nor the SFR-weighted or mass-weighted mock host distributions expected for the general star-forming galaxy population, instead occupying the extreme low-mass, high-sSFR, and low-metallicity end of host-galaxy parameter space.
We compare these hosts with well-defined samples of hydrogen-poor superluminous supernovae (SLSNe-I), hydrogen-rich superluminous supernovae (SLSNe-II), and Type II core-collapse supernovae (CCSNe).
Using cumulative distribution functions and statistical tests, we find that FRB-PRSs tend to reside in dwarf galaxies with high specific star formation rates and low gas-phase metallicities, even relative to galaxies of similar stellar mass. 
Their hosts are most similar to those of SLSNe-I, but differ significantly from CCSN hosts, with SLSN-II hosts lying between these populations.
These results suggest that FRB-PRSs do not simply trace star formation activity or the general FRB host population, but instead favor SLSN-I-like environments.
This preference is consistent with young magnetars formed through low-metallicity massive-star channels, for which SLSN-I-like ejecta imply a characteristic free--free transparency time of $\sim29$ yr at 600 MHz.

\end{abstract}

\keywords{Radio transient sources; Radio bursts; Supernovae; Dwarf galaxies}

\section{Introduction}
Fast radio bursts (FRBs) are bright, millisecond-duration radio transients first discovered by \citet{Lorimer2007}.
Since their discovery, a rapidly growing sample of FRBs has revealed that most FRBs have been detected as single-burst events, commonly referred to as apparently non-repeating FRBs, while approximately 100 FRBs have been observed to repeat over a wide range of timescales \citep{Chime2023,Cook2026}, with some sources even exhibiting extremely high burst rates \citep[e.g.,][]{Niu2022,Zhou2026}.
In recent years, the CHIME/FRB survey \citep{CHIME2018,Chime2026} has made a major contribution to detecting FRBs with its wide field and high cadence.
In addition, the high sensitivity of the Five-hundred-meter Aperture Spherical radio Telescope \citep[FAST;][]{Nan2011,Li2018} has also been crucial for detecting weak bursts \citep[e.g.,][]{Li2021,Niu2022}, whereas long-term monitoring with smaller telescopes has contributed to the detection of rare but bright bursts \citep[e.g.,][]{Ould2026}.
The host environments and physical origins of FRBs remain debated, but the detection of a FRB burst from the Galactic magnetar SGR J1935+2154 supports magnetars as one possible FRB progenitor channel \citep{Bochenek2020,CHIME2020}.

A particularly interesting subset of FRB sources is spatially associated with compact, luminous, nonthermal persistent radio sources (PRSs).
Their high radio luminosities and off-nuclear locations make it unlikely that they are powered by star formation or active galactic nuclei (AGN) activity.
The first such association was found for the repeating source FRB 20121102A, which exhibits an extreme and evolving rotation measure (RM), indicating a dense and highly magnetized environment local to the FRB source \citep{Chatterjee2017,Tendulkar2017,Marcote2017,Bassa2017,Michilli2018}.
A second FRB-PRS system, FRB 20190520B, not only exhibits an extreme and evolving RM, but also shows an unusually large host-galaxy dispersion measure (DM) contribution \citep{Niu2022,Chen2025b}.
More recently, FRB 20240114A \citep{Bruni2025,Chen2025a} and FRB 20190417A \citep{Ibik2024,Moroianu2026} have also been associated with compact PRSs and star-forming dwarf host galaxies, similar to FRB 20121102A and FRB 20190520B \citep[e.g.,][]{Feng2025}.

Previous case studies and qualitative comparisons have noticed that FRB-PRS hosts appear to share similar galaxy properties with hydrogen-poor superluminous supernovae (SLSNe-I), with both populations preferentially residing in low metallicity, high star formation rate (SFR) dwarf galaxies \citep[e.g.,][]{Eftekhari2017,Nicholl2017,Tendulkar2017,Bassa2017,Moroianu2026}.
SLSNe-I are generally associated with massive-star explosions in metal-poor environments, which may help produce rapidly rotating compact remnants \citep[e.g.,][]{Fruchter2006,Lunnan2014,Perley2016,Schulze2018,Taggart2021}.
However, it remains unclear whether the apparent similarity between FRB-PRS and SLSN-I hosts persists in a redshift-controlled population comparison, or whether FRB-PRS hosts simply represent the low-mass, high-SFR tail of the general star-forming galaxy or localized FRB hosts \citep{Bhandari2022}.

In this work, we compared the host-galaxy properties of FRB-PRSs with those of SLSNe-I, hydrogen-rich superluminous supernovae (SLSNe-II), Type II core-collapse supernovae (CCSNe), and FRBs without detected PRSs (non-PRS FRB) at $z<0.3$.
Using cumulative distribution functions and statistical tests widely used in comparative host-galaxy population studies \citep[e.g.,][]{Heintz2020,Bhandari2022,Gordon2023,Sharma2024}, we aim to determine whether FRB-PRS hosts are most similar to SLSN-I hosts among these comparison samples, thereby providing further evidence that FRB-PRSs preferentially occur in environments consistent with a young, massive-star or magnetar-related formation channel. 

\begin{table*}[htbp]
\centering
\tabcolsep=10pt
\caption{Host galaxy properties of known FRB-PRS systems.}
\label{tab:FRB-PRS properties}
\begin{tabular}{lcccc}
\hline\hline
Property & FRB 20121102A (1) & FRB 20190520B (2) & FRB 20240114A (3) & FRB 20190417A (4) \\
\hline
Redshift & 0.193 & 0.241 & 0.131 & 0.128 \\
$\log M_{*}$ ($\rm M_{\odot}$) & 8.1 $\pm$ 0.1 & 8.9 $\pm$ 0.1 & 8.6 $\pm$ 0.2 & 7.9 $\pm$ 0.1 \\
$\rm \log SFR_{H\alpha}$ ($\rm M_{\odot}\, yr^{-1}$) & -0.80 & -0.15 & -1.18 & -0.70 \\
$\rm \log sSFR_{H\alpha}$ ($\rm yr^{-1}$) & -8.90 & -9.05 & -9.78 & -8.60 \\
Gas-phase metallicity & 8.23 & 8.09 & 8.18 & 8.00 \\
\hline
\end{tabular}
\tablenotetext{}{\raggedright References:} (1) \citet{Bassa2017,Tendulkar2017}  (2) \citet{Chen2025b} (3) \citet{Bruni2025,Chen2025a} (4) \citet{Ibik2024,Moroianu2026}
\end{table*}

\section{Samples and Measurements} \label{Sec:Data}
We collected the host galaxy properties of all confirmed FRBs associated with compact PRSs, which are summarized in Table \ref{tab:FRB-PRS properties}.
The current sample includes FRB 20121102A, FRB 20190520B, FRB 20240114A, and FRB 20190417A, as well as a FRB-PRS candidate FRB 20201124A.
The latter remains controversial because its persistent emission is relatively weak \citep{Bruni2025} and spatially extended \citep{Piro2021,Ravi2022}, may be contaminated by star formation in its relatively massive host galaxy compared with other FRB-PRS hosts \citep[e.g.,][]{Dong2024}.
Throughout this work, our fiducial FRB-PRS sample therefore refers only to the four confirmed compact-PRS systems, but we have verified that including FRB 20201124A does not change our main results.
We derived the SFR from the extinction-corrected $\rm H\alpha$ luminosity using the relation $\rm \log(SFR_{H\alpha}[M_{\odot }\, yr^{-1}])=\log(5.5\times10^{-42}\, {\it L}_{H\alpha}\, [erg\,s^{-1}])$ from \citet{Kennicutt2012}.
We estimated gas-phase metallicities using the N2 calibration of \citet{Curti2020} (hereafter Curti20).
We also adopted the mass--metallicity relation (MZR) of Curti20 as the reference relation, and accounted for its redshift evolution by changing the turnover mass following \citet{Zahid2014}.

We constructed the SLSN comparison sample by collecting well-observed and measured host properties required for our analysis, particularly spectral observations.
For SLSNe-I, we mainly used the host sample compiled by \citet{Japelj2016}, which includes host galaxy properties from \citet{Lunnan2014}, \citet{Angus2016}, and \citet{Perley2016}, supplemented by the low-redshift host sample of \citet{Chen2017}.
We accounted for a $\sim$0.2 dex systematic difference in the stellar masses from \citet{Chen2017} to place them approximately on the same mass scale before combining the samples.
For SLSNe-II, we similarly adopted the spectroscopically well-observed host samples of \citet{Leloudas2015} and \citet{Perley2016}.
Although substantially larger SLSN-I and SLSN-II catalogs have been available recently \citep[e.g.,][]{Schulze2018,Taggart2021,Kangas2022,Gomez2024,Pessi2025}, these catalogs are primarily designed to study the transient population or do not provide homogeneous measurements of the host-galaxy properties, especially the optical emission-line measurements required to derive SFRs and gas-phase metallicities.
For comparison with typical massive-star explosions, we also used the Type II CCSN host sample of \citet{Schulze2021}.
Unlike the SLSN samples, the CCSN sample is used only as a reference population in the stellar mass and star formation comparisons and is not included in our gas-phase metallicity analyses.
Therefore, the CCSN sample does not require the same spectroscopic selection adopted for the SLSN hosts.
It is worth noting that, given the bursty star-formation histories of the host galaxies of FRB-PRSs, SLSNe-I, and SLSNe-II, their SFRs were measured from extinction-corrected nebular emission lines to trace recent star formation activity, whereas the SFRs of CCSN hosts were derived from SED fitting in \citet{Schulze2021}. 
Although the two indicators probe different star-formation timescales, studies of star-forming galaxies find no strong systematic offset between SED- and emission-line-based SFRs \citep[e.g.,][]{Salim2016} for star-forming galaxies. 
Therefore, the difference between the two SFR indicators is not expected to significantly affect our results.

We also collected host-galaxy properties for well-localized FRB host galaxies presented by \citet{Sharma2024} and generated the mock galaxy samples using v1.1.0 \texttt{GALFRB}\footnote{\url{https://github.com/loudasnick/GALFRB}} from \citet{Loudas2025}.
These samples were used as additional comparison populations to examine whether FRB-PRS hosts are representative of the general FRB host population or occupy a more specific region of host-galaxy parameter space.
Specifically, the observed FRB host sample includes 23 hosts localized by ASKAP \citep{Gordon2023}, 26 hosts from DSA-110 \citep{Sharma2024}, and 4 hosts from CHIME \citep{Bhardwaj2024}, which has been widely used in recent studies because of their relatively high completeness and homogeneous measurements \citep[e.g.,][]{Loudas2025,Horowicz2026}.
The \texttt{GALFRB} mock samples are drawn from a general galaxy population model, with stellar masses sampled from the two-component Schechter stellar mass function of \citet{Leja2020} and SFRs generated from the full probability density function $\rho({\rm SFR},\,M_{*},\, {\rm z})$ presented by \citet{Leja2022}.
The mock-galaxy and observed FRB-host properties were both derived using the same spectral energy distribution modeling framework, \texttt{Prospector} \citep{Leja2017,Johnson2021}, built by \citet{Leja2019}.
We adopted three mock galaxy samples with different weighting, including unweighted, SFR-weighted, and mass-weighted samples.
More details of \texttt{GALFRB} can be found in \citet{Loudas2025}, and a similar approach to generating mock samples is presented in \citet{Horowicz2026}.

To minimize potential systematic differences caused by galaxy evolution across redshift, we limit our analysis to sources at $z<0.3$ to match the redshift distribution of the FRB-PRS hosts.
The median redshifts of the redshift-controlled FRB-PRS, SLSN-I, SLSN-II, and non-PRS FRB host samples are 0.162, 0.169, 0.229, and 0.150, respectively, and their redshift distributions are statistically consistent with one another.
The mock galaxy samples are also generated over the same redshift interval $0<z<0.3$ to match the observational samples.
We have tested for possible redshift-dependent selection effects by examining the transient peak absolute magnitude and host stellar mass as functions of redshift.
We find no evidence for a strong redshift-dependent correlation in peak absolute magnitude and host stellar mass, with low correlation coefficients and large $p$-values.
Therefore, the residual luminosity incompleteness does not significantly affect our statistical comparisons of host-galaxy stellar mass.
By contrast, the CCSN host sample has a lower median redshift of 0.040, owing to the lower intrinsic luminosities of CCSNe compared with FRBs and SLSNe, which makes them preferentially detected at lower redshifts.
To further account for the potential bias introduced by the redshift evolution of SFR, which could affect the statistical comparisons, we also calculate the star-formation activity of each host galaxy relative to the star-forming main sequence, in addition to using specific SFR.
Specifically, we calculate the offset from the star-forming main sequence, $\Delta{\rm MS}$, for each galaxy at its corresponding redshift.
The detailed method and the corresponding statistical results are presented in Appendix \ref{appendix:MS}.

\section{Results}\label{Sec:Results}

\subsection{Comparison with SN hosts}
\begin{figure*}[htbp]
	\centering
	\includegraphics[width=0.9\linewidth]{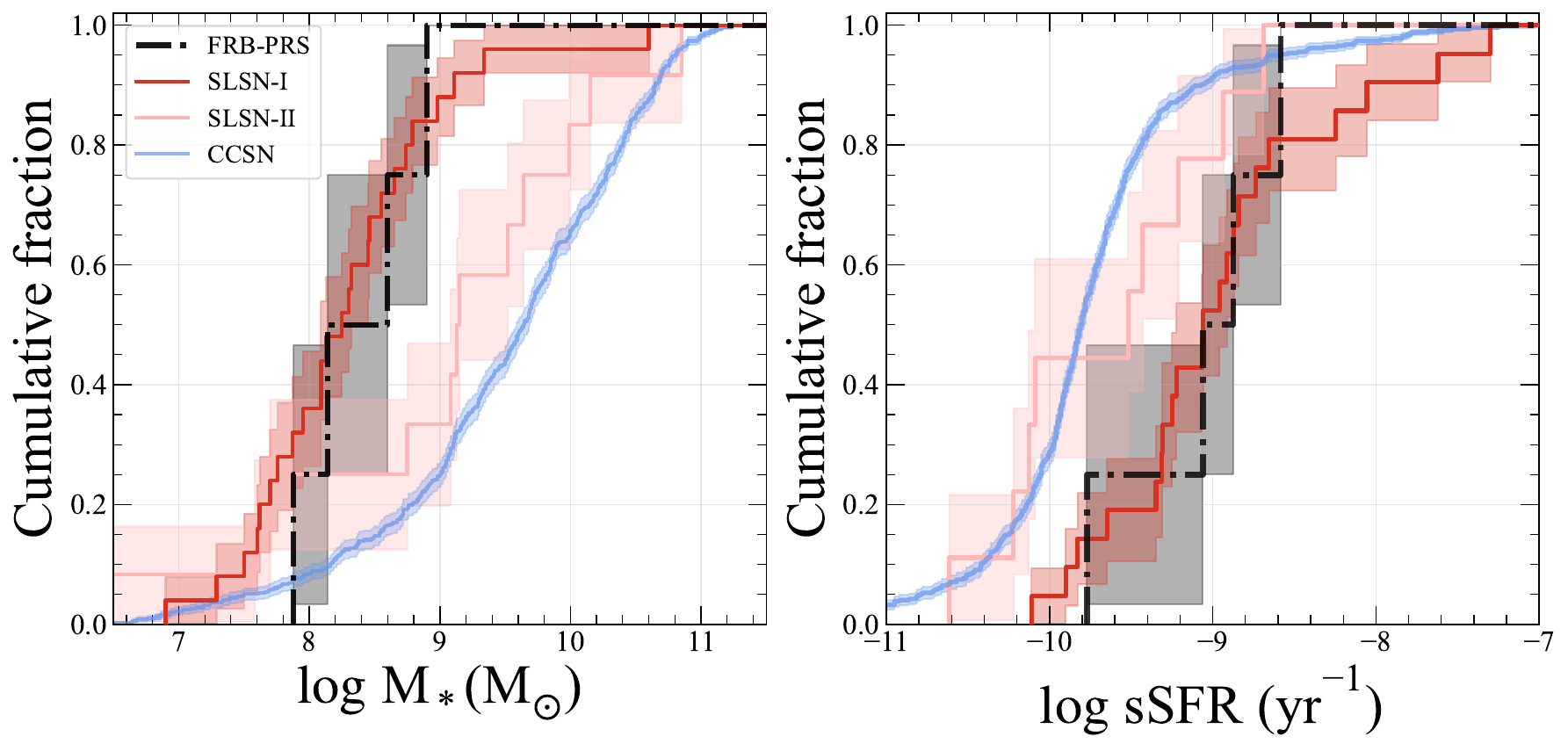}
	\caption{CDFs of stellar mass (left) and specific SFR (right) for FRB-PRS hosts (black), SLSN-I hosts (red), SLSN-II hosts (light red), and CCSN hosts (blue) in the $z<0.3$ redshift range. The shaded regions indicate the approximate $1\sigma$ binomial uncertainties of the CDFs.}
	\label{fig:Cumulative_fraction_M_sSFR.pdf}
\end{figure*}

\begin{figure*}[htbp]
	\centering
	\includegraphics[width=0.9\linewidth]{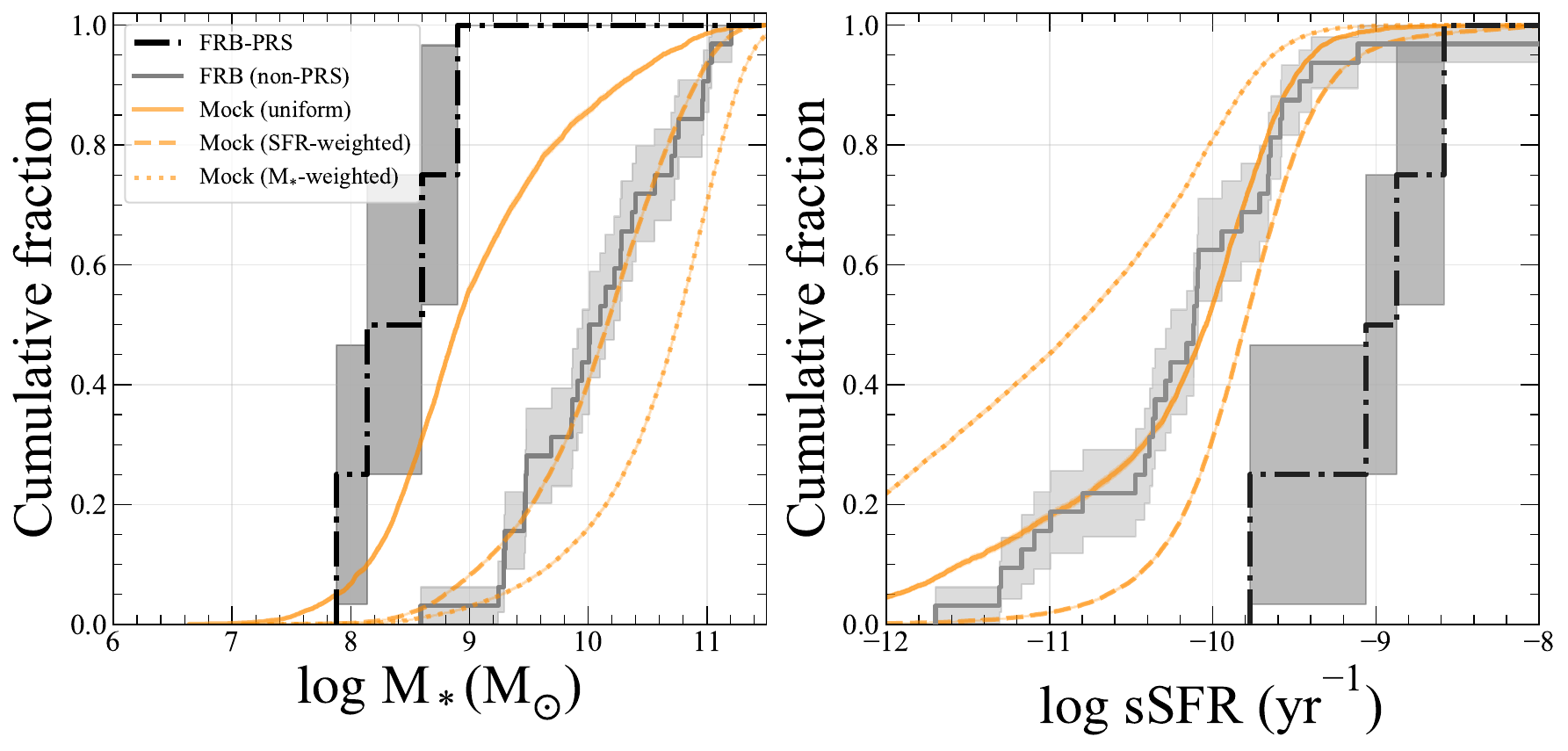}
	\caption{CDFs of stellar mass (left) and specific SFR (right) for FRB-PRS hosts (black), non-PRS FRB hosts (grey), uniform mock galaxy sample (orange solid curve), SFR-weighted mock galaxy sample (orange dashed curve), and $M_{*}$-weighted mock galaxy sample (orange dotted curve) in the $z<0.3$ redshift range.}
	\label{fig:Cumulative_fraction_FRB.pdf}
\end{figure*}

\begin{table*}[htbp]
\centering
\caption{Test statistics and p-values from the KS tests, AD tests, and bootstrap analysis for the stellar mass and sSFR distributions of SLSN-I, SLSN-II, CCSN, and non-PRS FRB host galaxies.}
\label{tab:KStest_bootstrap}
\tabcolsep=8pt
\renewcommand\arraystretch{1.1}
\hspace*{-2cm}
\begin{tabular}{lccc|ccc}
\hline\hline
Host property$^{a}$ & Comparison sample & Sample number & Median value & Method & Statistic & p-value \\
\hline
\multirow{4}{*}{Stellar mass (8.37)} 
& SLSN-I & 25 & 8.25 &\multirow{4}{*}{KS test}& 0.320 & 0.767 \\
& SLSN-II & 12 & 9.14 & & 0.667 & 0.112 \\
& CCSN & 496 & 9.65 & & 0.772 & 0.006 \\
& non-PRS FRB & 32 & 10.06 & & 0.969 & $<$0.001 \\
\hline
\multirow{4}{*}{Stellar mass (8.37)} 
& SLSN-I & 25 & 8.25 & \multirow{4}{*}{AD test} & -0.812 & 0.250 \\
& SLSN-II & 12 & 9.14 & & 1.049 & 0.121 \\
& CCSN & 496 & 9.65 & & 3.489 & 0.013 \\
& non-PRS FRB & 32 & 10.06 & & 7.399 & 0.001 \\
\hline
\multirow{4}{*}{Stellar mass (8.37)} 
& SLSN-I & 25 & 8.25 & \multirow{4}{*}{Bootstrap} & 0.120 & 0.699 \\
& SLSN-II & 12 & 9.14 & & -0.770 & 0.292 \\
& CCSN & 496 & 9.65 & & -1.275 & 0.044  \\
& non-PRS FRB & 32 & 10.06 & & -1.685 & $<$0.001 \\
\hline
\hline
\multirow{4}{*}{sSFR (-8.97)} 
& SLSN-I & 21 & -9.06 &\multirow{4}{*}{KS test}& 0.226 & 0.980 \\
& SLSN-II & 9 & -9.52 & & 0.528 & 0.336 \\
& CCSN & 496 & -9.81 & & 0.651 & 0.036  \\
& non-PRS FRB & 32 & -10.12 & & 0.719 & 0.027 \\
\hline
\multirow{4}{*}{sSFR (-8.97)} 
& SLSN-I & 21 & -9.06 &\multirow{4}{*}{AD test}& -1.123 & 0.250 \\
& SLSN-II & 9 & -9.52 & & 0.719 & 0.167 \\
& CCSN & 496 & -9.81 & & 2.991 & 0.020 \\
& non-PRS FRB & 32 & -10.12 & & 4.489 & 0.005 \\
\hline
\multirow{4}{*}{sSFR (-8.97)} 
& SLSN-I & 21 & -9.06 & \multirow{4}{*}{Bootstrap} & 0.091 & 0.788 \\
& SLSN-II & 9 & -9.52 & & 0.551 & 0.097 \\
& CCSN & 496 & -9.81 & & 0.838 & 0.020  \\
& non-PRS FRB & 32 & -10.12 & & 1.148 & $<$0.001 \\
\hline

\end{tabular}
\tablenotetext{a}{\raggedright The median values of the FRB-PRS sample are shown in parentheses.}
\end{table*}

In Figure \ref{fig:Cumulative_fraction_M_sSFR.pdf}, we compare the cumulative distribution functions (CDFs) of stellar mass ($\log\, M_{*}$, left panel) and specific SFR (sSFR, right panel) for the host galaxies of FRB-PRSs, SLSNe-I, SLSNe-II, and CCSNe.
As shown in the left panel, the FRB-PRS hosts are concentrated toward the low-mass end of the distribution, consistent with the SLSN-I hosts, but clearly offset from the CCSN hosts with higher stellar masses.
The right panel shows a similar trend in sSFR: FRB-PRS and SLSN-I hosts occupy a similar high-sSFR regime, with values clearly higher than those of CCSN hosts.
The SLSN-II hosts are shifted toward higher stellar masses and lower sSFRs than the FRB-PRS and SLSN-I hosts, but intermediate between these populations and the CCSN hosts.

We used two statistical methods together with a bootstrap analysis to quantify the potential similarities between the FRB-PRS hosts and the other host samples.
We performed two-sample Kolmogorov-Smirnov (KS) tests and Anderson-Darling (AD) tests to compute the statistics and associated $p$-values ($p_{\rm KS}$ and $p_{\rm AD}$) to test the null hypothesis that the two distributions are drawn from the same parent distribution.
Given the small number of FRB-PRS hosts, we further used a bootstrap analysis to verify the results from the KS and AD tests.
For each comparison sample, we randomly selected subsamples with replacement, each having the same size as the FRB-PRS sample.
We then calculated the median value of the considered host property for each subsample and repeated this procedure 10,000 times.
This procedure generates the distribution of median values, which can be compared with the observed median of the FRB-PRS hosts.
We defined $p$-value ($p_{\rm b}$) as twice the smaller of the fractions of resampled medians lying below or above the observed FRB-PRS median. 
Therefore, a small $p_{\rm b}$ indicates that the observed FRB-PRS median lies in one of the extreme tails of the resampled median distribution.
The results are summarized in Table \ref{tab:KStest_bootstrap}.

According to the three statistical methods, the FRB-PRS hosts are statistically distinct from the CCSN hosts in every parameter ($p$-values $<$0.05) and exhibit large values of the corresponding test statistic.
In contrast, SLSN-I hosts yield the smallest statistics and the smallest difference in median properties with FRB-PRS hosts.
Both the KS and bootstrap tests show large $p$-values ($>0.5$), indicating no statistically significant difference between the current FRB-PRS and SLSN-I host samples, and the AD test $p$-value is slightly lower ($p_{\rm AD}\sim 0.25$) but still above 0.05.
The corresponding test statistics are also the lowest among all comparisons.
SLSN-II hosts fall intermediately between SLSN-I and CCSN hosts but their $p$-values are also above 0.05, indicating that we cannot reject the null hypothesis that the FRB-PRS and SLSN-II hosts are drawn from the same parent distribution.

\subsection{Robustness test against missing massive hosts}
Because both FRB-PRS and SLSN host galaxy samples may be affected by selection effects, we explicitly test how many missing massive hosts would be required to erase the observed similarity. 
For FRB-PRS systems, the detectability of compact persistent radio emission depends on the radio background of the host galaxy, e.g., intrinsic PRS luminosity, source age and distance, observational sensitivity, and the radio continuum emission from the star formation and AGN of the host galaxy.
In particular, the compact persistent emission may be easier to identify in low-mass dwarf galaxies than in dusty and strongly star-forming massive galaxies where radio emission from star formation can be significant \citep{Sun2026}.
The radio emission from the PRSs could be covered or confused by the radio continuum emission from star-forming activities \citep[e.g., FRB 20201124A,][]{Dong2024,Bruni2025,Mfulwane2026} or AGN activity \citep{Law2022,Ibik2024}.
Similarly, SLSNe are primarily discovered in the optical band, making them more difficult to detect in dusty host galaxies \citep{Lunnan2014,Perley2016,Japelj2016}.
These effects may cause some sources located in massive galaxies to be missed.
Therefore, we performed a robustness test by artificially adding massive host galaxies to the FRB-PRS or SLSN-I host sample until the hypothesis that the two samples are drawn from the same distribution would be rejected.

First, we progressively added artificial host galaxies with $\log\, M_{*}/\rm{M}_{\odot}=10$, because galaxies around this stellar mass typically have the highest SFRs and lie near the turnover of the main sequence \citep{Lee2015}, which is expected to most strongly reduce the sample completeness of the FRB-PRS sample.
We find that at least 5 (4) additional sources are required to reduce $p_{\rm KS}$ ($p_{\rm AD}$) for the comparison between the FRB-PRS and SLSN-I hosts to $<0.05$.
Compared with the current sample of 4 sources, this result suggests that the number of additional massive hosts would have to exceed the size of the current FRB-PRS sample to significantly affect our statistical results.
This test shows that the present statistical result would not be affected by only a few missing massive hosts, although larger samples are needed to quantify this possibility more rigorously.
Similarly, from the perspective of potential incompleteness in the SLSN-I sample, at least 36 and 27 additional massive host galaxies would be needed to reduce $p_{\rm KS}$ and $p_{\rm AD}$ below 0.05, respectively, which are also more than one times the size of the parent sample.
This test shows that the observed similarity between FRB-PRS and SLSN-I hosts would not be erased by only a small number of missing massive host galaxies.
Meanwhile, the unusually high FRB activity of these systems may help mitigate confusion with radio emission from star formation and AGN by facilitating precise localization and targeted high-resolution radio follow-up. 
In addition, observations of known repeating FRBs suggest that luminous PRSs are uncommon \citep{Ibik2024}.

\subsection{Comparison with FRB hosts and mock galaxy samples}
Figure \ref{fig:Cumulative_fraction_FRB.pdf} displays the CDFs of $\log\, M_{*}$ and sSFR for FRB-PRS, non-PRS FRB, and three mock samples: uniform (unweighted), SFR-weighted, and mass-weighted samples.
For the CDF of stellar mass in the left panel, the SFR-weighted population matches the observed distribution of non-PRS FRB hosts very well within the redshift range $z<0.3$, which is consistent with the results from \citet{Loudas2025}.
The FRB-PRS hosts are visibly shifted relative to the non-PRS FRBs and all three mock galaxy samples, because of the lack of massive galaxies in the FRB-PRS sample.
Furthermore, FRB-PRS hosts also favor systematically higher sSFR values than those of other samples shown in the right panel, suggesting a distinct host-galaxy environment for FRB-PRSs.
We quantified the differences between FRB-PRS and non-PRS FRB hosts with all three statistical methods, summarized in Table \ref{tab:KStest_bootstrap}.
FRB-PRS hosts are statistically distinct from the non-PRS FRB hosts in both the stellar mass ($p_{\rm KS}<0.001$, $p_{\rm AD}=0.001$, and $p_{\rm b}<0.001$) and sSFR ($p_{\rm KS}=0.027$, $p_{\rm AD}=0.005$, and $p_{\rm b}<0.001$).
Together with their offsets from the mock galaxy samples, these results suggest that the currently identified FRB-PRS hosts occupy a special region of host-galaxy parameter space that is not representative of the FRB host population or the general star-forming galaxy population.

\subsection{Spectroscopic analysis and metallicity}
\begin{figure*}[htbp]
	\centering
	\includegraphics[width=1\linewidth]{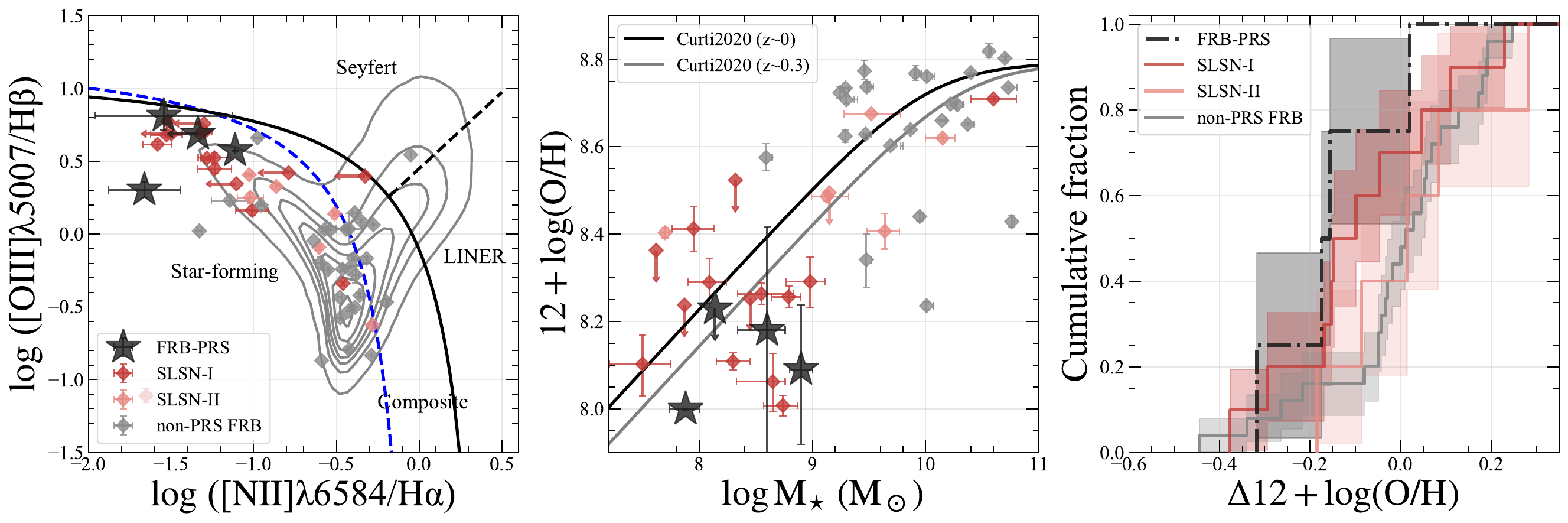}
	\caption{Left: BPT diagram for the host galaxies of FRB-PRS (black stars), SLSN-I (red diamonds), SLSN-II (light red diamonds), and non-PRS FRB (grey diamonds). The blue dashed line, black solid line, and black dashed line indicate the division between star-forming galaxies and composite galaxies \citep{Kauffmann2003}, composite sources and AGN \citep{Kewley2001}, and Seyfert and LINER galaxies \citep{Schawinski2007}. The contours indicate the SDSS galaxies from the MPA-JHU catalog \citep{Kauffmann2003,Brinchmann2004}. Middle: Comparing the FRB-PRS, SLSN-I, SLSN-II, and non-PRS FRB host galaxies with the mass--metallicity relation, overplotting the local relation from \citet{Curti2020} and a $z\sim0.3$ version in which the turnover mass evolves following \citet{Zahid2014}. Right: CDFs of $\rm \Delta\,[12+\log(O/H)]$, defined as the offset between the observed gas-phase metallicity and the expected value from the MZR at the stellar mass and redshift of each galaxy.}
	\label{fig:Metallicity_N2.pdf}
\end{figure*}

We then investigated the ionization properties of FRB-PRS host galaxies in terms of the $\OiiiHb$ and $\NiiHa$ emission-line ratios and gas-phase metallicity, and compared them with those of SLSN-I, SLSN-II, and non-PRS FRB hosts.
The left panel of Figure \ref{fig:Metallicity_N2.pdf} shows the Baldwin, Phillips, and Terlevich diagram \citep[BPT;][]{Baldwin1981} for the host galaxies of the four samples, overlaid on the distribution of SDSS galaxies from the MPA-JHU catalog \citep{Kauffmann2003,Brinchmann2004}.
All FRB-PRS hosts and most SLSN-I hosts lie in the upper-left region of the BPT diagram, consistent with low-metallicity, highly-ionized systems characterized by high $\OiiiHb$ and low $\NiiHa$ emission-line ratios.
Most SLSN-II and non-PRS FRB hosts are classified as normal star-forming galaxies, while a small fraction fall into the composite region, and only one non-PRS FRB host lies in the Seyfert region.

The MZR reflects the chemical enrichment history of galaxies and plays a key role in constraining galaxy evolution processes \citep[e.g.,][]{Lequeux1979,Andrews2013}.
We compared the distributions of the four samples in the MZR, together with the empirical relation from \citet{Curti2020}, as shown in the middle panel of Figure \ref{fig:Metallicity_N2.pdf}.
All FRB-PRS and most SLSN-I hosts lie below the reference MZR, suggesting that these systems have lower gas-phase metallicities than typical galaxies of similar stellar mass.
SLSN-II and non-PRS FRB hosts are located toward the high stellar mass end and broadly follow the MZR, except for a few galaxies that exhibit relatively low metallicities.

We then calculated $\rm \Delta\,[12+\log(O/H)]$, which is defined as the offset between the observed gas-phase metallicity and the MZR at the corresponding stellar mass and redshift of each source, and compared the CDFs of $\rm \Delta\,[12+\log(O/H)]$ for the four samples.
The results are shown in the right panel of Figure \ref{fig:Metallicity_N2.pdf}.
FRB-PRS and SLSN-I hosts have visually similar CDFs of $\rm \Delta\,[12+\log(O/H)]$, whereas SLSN-II and non-PRS FRB hosts are visually offset from them.
The results suggest that the low metallicities of FRB-PRS and SLSN-I hosts are not simply due to their low stellar masses, but that they preferentially occupy the more metal-poor end of the dwarf-galaxy population.
Furthermore, because $\Nii$ becomes weaker at lower metallicity, any selection bias based on a measurable $\Nii$ line would preferentially miss the most metal-poor systems, rather than produce the observed negative MZR offsets.

\subsection{Two-dimensional distribution of star formation and metallicity}
\begin{figure}[htbp]
	\centering
	\includegraphics[width=1\linewidth]{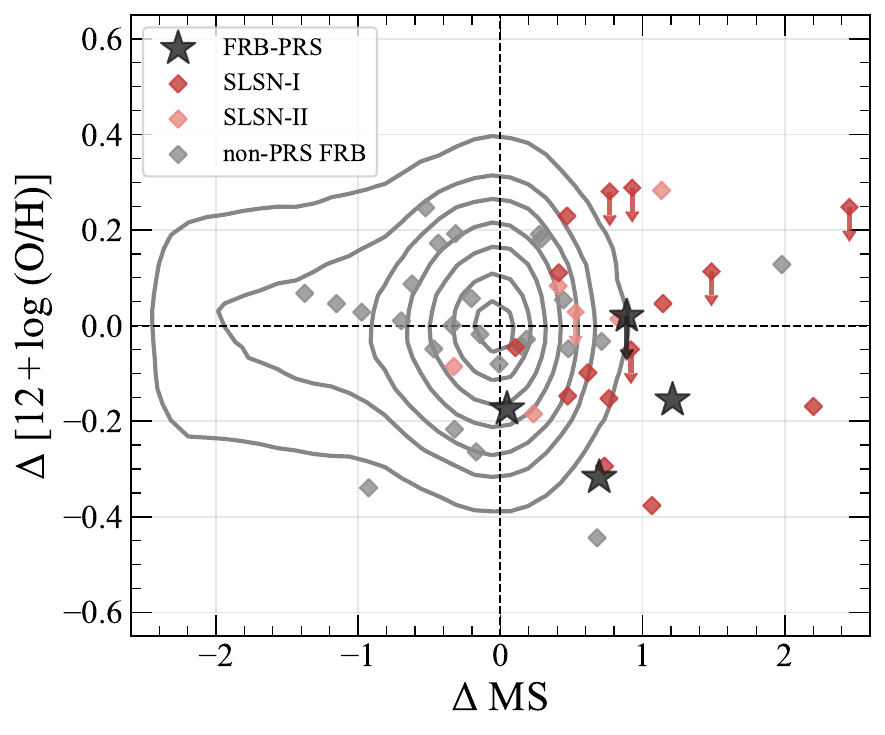}
	\caption{Comparison of the FRB-PRS, SLSN-I, SLSN-II, and non-PRS FRB host galaxies in the two-dimensional plane of $\Delta\, [12+\log({\rm O/H})]$ versus $\Delta\,{\rm MS}$. The contours represent the unweighted mock galaxy sample, whose gas-phase metallicities were randomly sampled from the MZR of \citet{Curti2020} assuming a 0.15 dex scatter.
}
	\label{fig:2D_SFR_Metallicity.pdf}
\end{figure}

We then measured the offsets of FRB-PRS, SLSN-I, SLSN-II, and non-PRS FRB host samples relative to the star-forming main sequence (MS) and the MZR, to construct the two-dimensional distribution of $\Delta\, [12+\log({\rm O/H})]$ versus $\Delta\,{\rm MS}$, overlaid with the mock unweighted galaxy sample in Figure \ref{fig:2D_SFR_Metallicity.pdf}.
Compared with the mock galaxies, the FRB-PRS hosts tend to occupy the quadrant with enhanced SFRs relative to the MS and low metallicities relative to the MZR.
Most SLSN-I hosts show a similar trend \citep{Perley2016,Schulze2018,Taggart2021}, although a few exhibit relatively high gas-phase metallicities, whereas SLSN-II and non-PRS FRB hosts do not exhibit a clear preference in this two-dimensional parameter space.
In particular, the distribution of non-PRS FRB hosts is consistent with the mock galaxy contours. 
This two-dimensional distribution strengthens the conclusion that FRB-PRSs favor SLSN-I-like environments.

\section{Discussion}\label{Sec:Discussion}

Some previous studies of FRB-PRS host galaxies have noticed, largely from individual case studies \citep[e.g.,][]{Tendulkar2017,Bassa2017,Niu2022,Bruni2025} and diagnostic diagrams \citep[e.g.,][]{Chen2025b,Moroianu2026}, that they may occupy host-galaxy parameter space similar to that of SLSN hosts.
In this work, we turn this qualitative impression into statistical tests with several redshift-controlled and well-defined host-galaxy samples.
Among the comparison samples considered, FRB-PRS hosts occupy a region of parameter space most similar to that of SLSN-I hosts, but differ from CCSN hosts, with SLSN-II hosts lying between the two populations.
The FRB-PRS and SLSN-I hosts show similar distributions not only in stellar mass and sSFR, but also in gas-phase metallicity, suggesting that their progenitors may have similar environmental preferences.
This result indicates that FRB-PRSs are unlikely to simply trace the star formation activity and general population of massive-star core-collapse events or the bulk of localized FRBs, and instead favor a more restricted environment similar to that of SLSNe-I.
They may require more specific environmental or progenitor conditions, such as low metallicity, which can reduce stellar-wind mass loss and retain sufficient angular-momentum to form rapidly rotating magnetars \citep[e.g.,][]{Woosley2006}.
Although the similarity between FRB-PRS and SLSN-I host galaxies does not prove that the two populations share the same progenitors, current models seem to favor magnetars as a common central engine.
SLSNe-I are generally interpreted as the core collapse of massive stars in low-metallicity environments, leading to the formation of rapidly rotating magnetars whose spin-down energy powers the luminous supernova emission \citep{Kasen2010,Woosley2010,Perley2016,Nicholl2017a,Gal-Yam2019}.
For FRB-PRS, a magnetar wind nebula \citep[e.g.,][]{Metzger2017,Beloborodov2017,Margalit2018,Zhao2021} or a young neutron star embedded in a supernova remnant \citep{Connor2016} can provide a possible explanation for the associated compact PRS and the large DM and rapidly variable RM observed in some systems \citep[e.g.,][]{Michilli2018,Niu2022,Niu2026}.
Therefore, the similarity between FRB-PRS and SLSN-I hosts supports the possibility that at least some FRB-PRSs are linked to young magnetars formed through massive-star channels in metal-poor, actively star-forming environments.
In this scenario, some FRB-PRS systems may even represent young magnetar remnants produced by SLSN-I-like explosions on timescales of years to decades \citep[e.g.,][]{Metzger2017,Nicholl2017,Omand2018,Eftekhari2019}.
We note that larger FRB-PRS samples and direct detections of FRB-PRS systems associated with SLSNe-I are needed to improve the connection between FRB-PRSs and SLSNe-I in the future.

We estimated a characteristic transparency time for SLSN-I using the analytic free--free transparency scaling adopted by \citet{Dong2025}, based on the treatment of ionized SN ejecta in \citet{Metzger2017}, and primarily determined by the ejecta mass and expansion velocity \citep{Murase2016,Piro2016}:
\begin{equation}
\begin{aligned}
t^{\rm ff}_{\nu} \simeq {}& 6~{\rm yr}\,
(1+z)^{3/5}
\left(\frac{\nu}{600~{\rm MHz}}\right)^{-2/5}
\left(\frac{M_{\rm ej}}{0.6\,M_\odot}\right)^{2/5} \\
&\times
\left(\frac{v_{\rm ej}}{10^4~{\rm km\,s^{-1}}}\right)^{-1}.
\end{aligned}
\end{equation}
We adopted the mean ejecta mass ($M_{\rm ej}=9.3^{+12.9}_{-4.8}\ M_{\odot}$) and velocity ($v_{\rm ej}=6,800^{+3,400}_{-2,000}{\rm \ km\ s^{-1}}$) inferred for the SLSN-I population by \citet{Gomez2024}.
At 600 MHz, the central frequency of CHIME, and adopting the median redshift of our SLSN-I sample ($z=0.169$), we obtained a characteristic free--free transparency time of $t^{\rm ff}_{\rm 600\,MHz}\approx 29.0^{+29.2}_{-14.5}$ yr. 
This is significantly longer than the minimum timescale of $\sim 6.4$ yr estimated by \citet{Dong2025} using the low ejecta mass ($M_{\rm ej}=0.62\ M_{\odot}$) and high ejecta velocity ($v_{\rm ej}=10,880{\rm \ km\ s^{-1}}$) of a strongly stripped CCSN.
If the systematically lower gas-phase metallicity of the FRB-PRS and SLSN-I hosts is also taken into account, the transparency time could become even longer, because lower metallicity reduces line-driven stellar-wind mass loss and may allow the SLSN-I progenitor to retain more mass before explosion \citep[e.g.,][]{Vink2001,Mokiem2007}.
It is worth noting that the actual transparency time also depends on the ionization state, temperature, and density structure of the expanding ejecta, so the value above should be interpreted as a characteristic timescale rather than a strict lower limit.
Overall, in this situation, even the oldest observed SLSN-I-like remnants may still not have fully reached, or may have only just begun to reach, the transparency window that CHIME can observe.
Since free--free absorption decreases toward higher frequencies, sensitive searches at higher radio frequencies, including with FAST, provide a promising way to detect FRB activity from young SLSN-I-like remnants and test whether FRB-PRS systems can arise from SLSN-I-like explosions.

\section{Conclusions}\label{Sec:Conclusion}
In this work, we compared the four confirmed FRB-PRS hosts with the host sample of SLSNe-I, SLSNe-II, CCSNe, and non-PRS FRBs at $z < 0.3$ to look for potential similarities or differences between these populations.
\begin{enumerate}
\item FRB-PRS hosts are broadly consistent with SLSN-I hosts in stellar mass, sSFR, and gas-phase metallicity, suggesting that their progenitors may have similar environmental preferences.
\item FRB-PRS hosts are clearly offset from both CCSN hosts and the SFR-weighted mock galaxy sample. 
The offset suggests that FRB-PRSs do not simply trace the star formation activity or the general population of massive-star core-collapse events.
\item FRB-PRSs appear to represent a special subset of FRBs. 
In addition to their high burst activity and associated PRSs, FRB-PRS hosts differ from non-PRS FRB hosts, occupying an extreme low-mass, high-sSFR, and low-metallicity region of the host-galaxy parameter space.
This distinction suggests that FRB-PRSs may represent a younger subpopulation of FRBs or a specific evolutionary stage in the lifetime of FRB progenitors.
\item Our results support a scenario in which some FRB-PRSs are linked to young magnetars formed through massive-star channels in metal-poor, actively star-forming environments, but larger samples are required to establish a direct progenitor connection. 
We estimate the characteristic free--free transparency time of SLSN-I ejecta to be $\sim 29$ yr at 600 MHz, implying that even the earliest discovered SLSNe-I may not yet have entered, or may only now be entering, the observational window for FRBs.

\end{enumerate}

\begin{acknowledgments}
This work was supported by the National Natural Science Foundation of China grant No. 12588202.
This work is partially supported by Guizhou Leading Innovative Talent Workstation for Extreme Universe.
DL acknowledges support from the New Cornerstone foundation.
\end{acknowledgments}

\twocolumngrid
\appendix
\counterwithin{figure}{section}
\renewcommand{\thefigure}{\thesection.\arabic{figure}}
\section{Cumulative distribution of star-formation main-sequence offset} \label{appendix:MS}

\begin{figure}[htbp]
	\centering
	\includegraphics[width=0.9\linewidth]{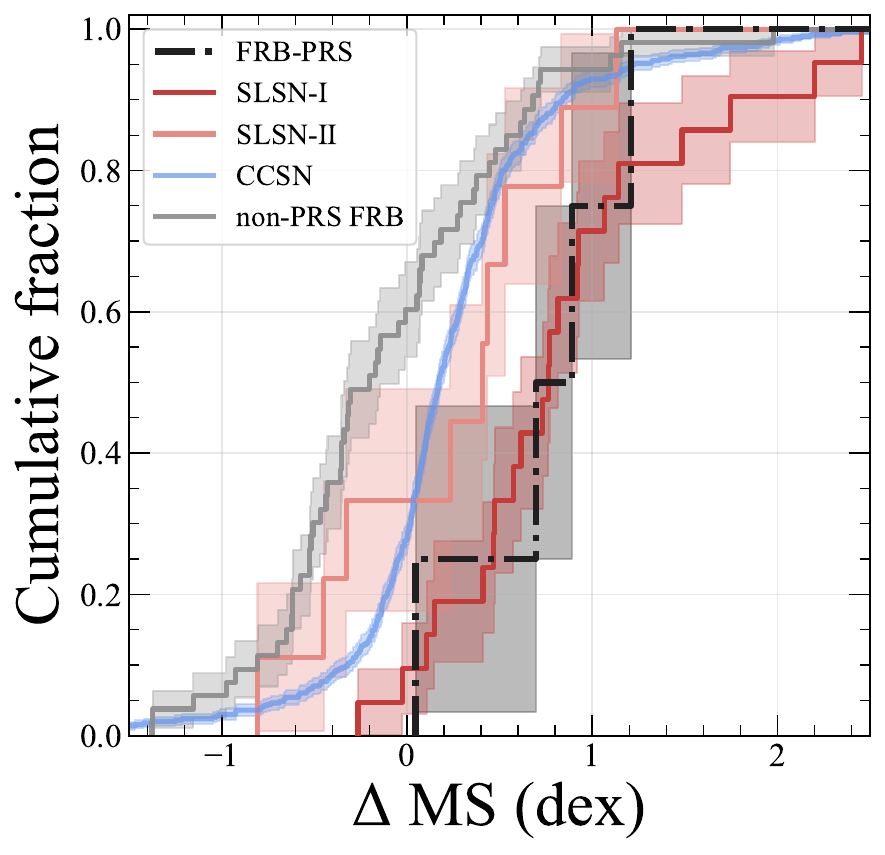}
	\caption{CDFs of $\Delta\ {\rm MS}$ for FRB-PRS hosts (black), SLSN-I hosts (red), SLSN-II hosts (light red), CCSN hosts (blue), and non-PRS FRB (grey) in the $z<0.3$ redshift range. The shaded regions indicate the approximate $1\sigma$ binomial uncertainties of the CDFs.}
	\label{fig:Cumulative_fraction_deltaMS.pdf}
\end{figure}

To account for the redshift evolution of star-formation activity, we further compare the CDFs of the offset from the star-forming main sequence, $\Delta{\rm MS}$, for each host galaxy.
We adopt the star-forming main sequence relation of \citet{Leja2022} and the offset is defined by the following equation.
\begin{equation}
\Delta\ {\rm MS} = \log_{10}{\rm SFR}-\log_{10}{\rm SFR_{MS}}(M_{*},z),
\end{equation}
where ${\rm SFR_{MS}}(M_{*},z)$ is the expected SFR of a galaxy on the star-forming main sequence at the corresponding stellar mass and redshift.
The resulting CDFs are shown in Figure \ref{fig:Cumulative_fraction_deltaMS.pdf}.
Compared with the right panel of Figure \ref{fig:Cumulative_fraction_M_sSFR.pdf}, the CCSN host distribution shifts slightly closer to that of the FRB-PRS hosts in $\Delta{\rm MS}$.
However, the two distributions remain significantly different, indicating that the enhanced star-formation activity of FRB-PRS hosts cannot be explained solely by the redshift evolution.

\bibliography{sample701}{}
\bibliographystyle{aasjournalv7}

\end{document}